\documentclass[twocolumn]{biophys-new}
\usepackage[utf8]{inputenc}
\usepackage{graphicx}
\usepackage[colorlinks,allcolors=cyan!70!black]{hyperref}

\usepackage{xcolor}
\definecolor{YKB}{rgb}{0.05,0.18,0.75}

\usepackage{lipsum}

\title{A topological look into the evolution of developmental programs}

\runningtitle{Topological look into development} %% For page header

\author[1,*]{Somya Mani}
\author[1,2,*]{Tsvi Tlusty}
\runningauthor{Mani and Tlusty} %% For page header

\affil[1]{Center for Soft and Living Matter, Institute for Basic Science, Ulsan 44919, Republic of Korea}
\affil[2]{Departments of Physics and Chemistry, UNIST, Ulsan 44919, Republic of Korea}

\corrauthor[*]{SM: somyamn@gmail.com, TT: tsvitlusty@gmail.com}

\papertype{Article}

\begin{document}

\begin{frontmatter}

\begin{abstract}
Rapid advance of experimental techniques provides an unprecedented in-depth view into complex developmental processes. Still, little is known on how the complexity of multicellular organisms evolved by elaborating developmental programs and inventing new cell types.  A hurdle to understanding developmental evolution is the  difficulty of even describing the intertwined network of spatiotemporal processes underlying the development of complex multicellular organisms. Nonetheless, an overview of developmental trajectories can be obtained from cell type lineage maps. Here, we propose that these lineage maps can also reveal how developmental programs evolve: the modes of evolving new cell types in an organism should be visible in its developmental trajectories, and therefore in the geometry of its cell type lineage map. This idea is demonstrated using a parsimonious generative model of developmental programs, which allows us to reliably survey the universe of all possible programs and examine their topological features. We find that, contrary to belief, tree-like lineage maps are rare and lineage maps of complex multicellular organisms are likely to be directed acyclic graphs where multiple developmental routes can converge on the same cell type. While cell type evolution prescribes what developmental programs come into existence, natural selection prunes those programs which produce low-functioning organisms. Our model indicates that additionally, lineage map topologies are correlated with such a functional property: the ability of organisms to regenerate. 
\end{abstract}

\begin{sigstatement}
Cell type invention is a chief process in the evolution of developmental programs. Traditionally, developmental trajectories are represented as cell type lineage maps. Here we propose that systematic analysis of these maps, in particular their topology, should reveal traces of the manner in which cell types were invented. This is illustrated using a generative model of developmental programs, which allows one to robustly survey the geometry of cell lineage maps and link them to modes of cell type invention.
We suggest that predictions made by such mathematical models, in conjunction with surveys of real cell-lineage maps of different multicellular lineages could uncover mechanisms underlying evolution of developmental programs.
\end{sigstatement}
\end{frontmatter}

\section*{Statistics of lineage maps reflect developmental evolution}

How can one understand the astounding richness of life forms?---While molecules and mechanisms of biological development are conserved within each multicellular lineage~\cite{meyerowitz2002plants}, these lineages are extraordinarily diverse; land plants and animals include many thousands to millions of species~\cite{knoll2011multiple}. This diversity is in part due to the distinct cell types present in different organisms. And in this sense, \hyperref[glossary]{developmental programs} evolve by inventing new cell types~\cite{arendt2016origin}. While ancestral lineages likely resembled the simplest multicellular organisms alive today, such as \textit{Volvox carterii} that has two cell types \cite{matt2016volvox}, the extant diversity of today's organisms ranges from those with a few cell types to those with hundreds. What molecular mechanisms and logic could produce such diversity remains a persistent question in development. 

One way to tackle this question is by comparing developmental programs across species of various levels of \hyperref[glossary]{complexity}. Analyzing developmental genes in order to see how gene families have expanded is fairly easy. But we now realize that this is far from sufficient, because genes interact combinatorially to express cell types. For example, looking at genomes of sponges, one might be tempted to conclude that they posses neurons since they have all the necessary components to make synapses. In reality, however, it was only in the bilaterian/cnidarian ancestor that these components were arranged in a manner that expresses the synapse \cite{arendt2016origin}. 

This example is a reminder that developmental programs are \textit{functions} or \textit{algorithms} for the assembly of organisms. Long ago it was recognized by Cantor that there are always many more conceivable functions than combinations of variables \cite{Cantor1891}. Perhaps the simplest example---which is used as a minimalist model of development \cite{albert2003topology}---is a Boolean function of $N$ binary variables. There are $2^N$ combinations of variables, but a much larger number, $2^{2^N}$, of possible Boolean functions  
\footnote{For finite sets, such as the $N$ binary variables, the result is proved very simply by enumerating all possible functions. The groundbreaking discovery of Cantor was that the result applies also to infinite sets.}. Cantor's theorem has two implications relevant to our discussion:

(I) First, one cannot infer a function merely by inspecting its variables . Or equivalently, listing the lines of a computer program without knowing how they are logically linked will not suffice to understand the algorithm. Thus, to describe development in any organism, it is essential to look at developmental genes in the context of their regulatory architecture. 

(II) However, as the complexity of the organism increases, this task quickly becomes infeasible because the number of possible functions or programs increases very steeply with the number components. This number  grows super-exponentially, like the number $2^{2^N}$ of Boolean functions 
\footnote{For example, $N=10$ binary variables have $2^{10}=1024$ combinations, while the number of Boolean functions of $10$ variables is $2^{2^{10}} \simeq 10^{308}$, much more than the number of particles in the universe.}. 
Thus, for very simple organisms, such as volvox with its two cell types, one may outline a complete picture of its developmental process~\cite{matt2016volvox}. But in complex organisms, such as humans with their 200 cell types \cite{milo2010bionumbers}, development is an elaborate process that involves coordinated gene expression, cell-cell communication, asymmetric cell division, cell movements, and cell death~\cite{barresi2020developmental}.

Such combinatorially-exploding complexity is daunting.  
Still, we do have a succinct, accessible representation of developmental events in terms of \textit{cell type lineage maps} (\hyperref[glossary]{CTLMs}), which detail how different cell types of the body are generated through cellular differentiation. Besides cataloguing differentiation events, CTLMs provide a glimpse into the underlying regulatory architecture. Implicit in these maps is information about which cell states are stable, and thus can be called cell types, into how many cell types any given type can differentiate, and how differentiation depends on \hyperref[glossary]{cellular context}. 

We propose here that CTLMs may also teach us about the evolution of developmental programs. Invention of a new cell type involves rewiring of the underlying regulatory architecture of development. This architecture controls not only the identities of cell types, but also the developmental \textit{trajectories} through which they are produced. Therefore, it is reasonable to assume that the mechanisms which lead to cell type invention should leave a mark on developmental trajectories, and on the geometry of lineage maps which essentially trace these trajectories. Indeed, in \cite{yuan2020alignment}, through a comparison of cell lineage maps of two nematode species, the authors were able to identify evolutionary correspondence between their cell types. Here we further suggest that the statistics of the \hyperref[glossary]{\textit{topology}} of lineage maps should reflect the modes through which cell types evolve. 

In this article, we first describe the role CTLMs play in the study of biological development, and dwell on prevalent biases in our conception of CTLMs. We then consider the role CTLMs could play in elucidating developmental evolution, and discuss this idea using a simple generative model \cite{mani2021comprehensive}. The model allows us to anticipate the statistical distribution of lineage graphs generated by distinct modes of cell type evolution. We examine a simple case (``null hypothesis") where developmental programs are not biased by any particular mode of cell type invention. Contrary to ingrained belief, our model shows that CTLMs are highly unlikely to be tree-like. Instead, they are likely to be directed acyclic graphs where a single cell type is reachable via multiple developmental routes.

The model also demonstrates how topologies of CTLMs can encode information about functional attributes of organisms. For instance, we see that the ability of organisms to regenerate is correlated with lineage map topology. Such correlations can form a basis for natural selection to favor certain developmental programs, and thereby reveal more layers in the multilevel process of developmental evolution.

\section*{Usefulness of cell type lineage maps in the study of development}

Bodies of multicellular organisms are composed of multiple types of cells, differing by the distinct functions they perform. For example, humans with their  $ \geq 200$ cell types \cite{milo2010bionumbers}, are composed of cells such as neurons that process and relay information, muscle cells that allow locomotion, B-cells that provide immunity, etc.

Irrespective of its complexity, any adult multicellular body is ultimately derived from a single celled zygote through the process of development. Historically, to trace development, embryonic cells were labelled with dyes which allowed following their divisions to identify the adult tissues they form. This mapping of embryonic cells to cell types in the adult is called fate mapping. The construction of fate maps provided insights into the mechanism of fate determination during development. Certain embryonic cells are \textit{autonomously specified}, meaning that their fates remain unchanged even when they are grown separately. For example, in the tunicate \textit{Styela partitia}, any separated embryonic cell type gives rise to the same adult cell types it normally does, and therefore yields partial adults. Other embryonic cells are \textit{conditionally specified}, for example in sea urchins, where separated embryonic cells behave differently, and develop into complete adults \cite{barresi2020developmental}.

Such mappings of cells produced by following sequential cell divisions are still useful today as a means to understand and represent developmental mechanisms. For example, in female \textit{Drosophila}, the cell divisions that give rise to the oocyte and associated nurse cells are highly regular, and yield a characteristic lineage tree. The closeness of nurse cells to oocytes in this lineage tree strongly determines their cell size \cite{alsous2017collective}.

\begin{figure*}[htb!]
\centering
\includegraphics[width=0.8\textwidth]{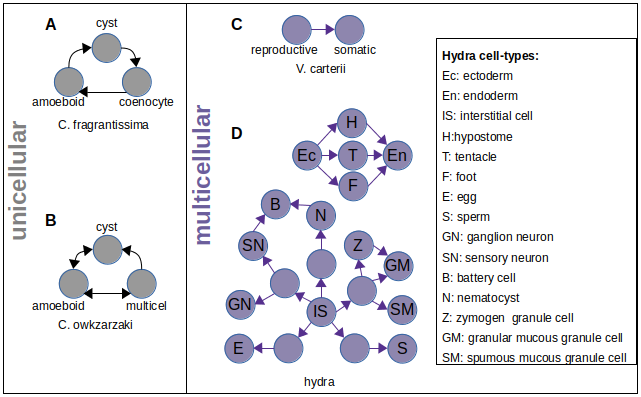}
\caption{Cell type lineage maps (CTLMs). In (A,B), gray circles represent cell-states of unicellular organisms, and arrows represent changes in cellular phenotype. In (C,D) the blue circles represent cell types of multicellular organisms, and arrows represent differentiation. (A) The life-cycle of the unicellular organism \textit{Creolimax fragrantissima} involves cycling between three stages: a motile amoeboid state, the immobile cyst state, and the multinucleate coenocyte. (B) \textit{Capsaspora owkzarzaki} responds to environmental cues to switch between three cell-states: cells switch from an reproductive amoeboid state to an \hyperref[glossary]{aggregative multicellular} state in the presence of nutrients, and both the amoeboid and multicellular cell-states switch to a cyst state under starvation. (C) The CTLM for both embryonic development, and adult homeostasis of \textit{Volvox carterii} \cite{matt2016volvox}. (D) the CTLM representing adult homeostasis in hydra \cite{siebert2019stem}. Empty circles represent unannotated intermediate cell types.
}
\label{fig1}
\end{figure*}

Molecular markers allow identification of cell types much more accurately and thereby facilitate refinement of fate maps. The most refined versions of fate maps are called \textit{cell type lineage maps} (CTLMs), and these track every single cellular differentiation event between the initial embryonic cells and the final adult cells. Mathematically speaking, a CTLM is a graph whose nodes represent distinct cell types, and directed edges represent differentiation of one cell type into another (Fig.\ref{fig1}). 

%%%%%%%%%%%%%%%%

\subsection*{Cell type lineage maps capture progression through life stages}
Generally, CTLMs can represent not only embryonic development, but other life-stages as well, such as adult homeostasis~\cite{plass2018cell}, metamorphosis~\cite{sogabe2016ontogeny} and regeneration~\cite{siebert2008cell}. In reality, especially for asexually reproducing organisms, it is not always possible to tell apart the CTLMs of different life-stages. For example, Fig.\ref{fig1}C can be said to represent both the embryonic development and the adult homeostatic map for volvox~\cite{matt2016volvox}, and while Fig.\ref{fig1}D represents the CTLM for adult homeostatis in hydra \cite{siebert2019stem}, it contains edges representing cellular differentiation that are also observed during hydra regeneration \cite{siebert2008cell}. That is, these organisms re-use the same differentiation pathways, and a single CTLM can sufficiently describe various life-stages. In contrast, developmental programs can also display extreme plasticity, for example, under unfavorable conditions, the immortal jellyfish \textit{T. dohrnii} can reverse its development \cite{matsumoto2019transcriptome}, essentially reversing edges across its CTLM. 

Fundamentally, CTLMs of various life stages of an organism reflect the parts of its `regulatory architecture'---the inter-cellular signaling system, and the gene regulatory network---that are accessed during these different stages in the organism's lifetime. And the plasticity of CTLMs indicates that multiple developmental routes can potentially be used to access the same cell type. Which developmental route is eventually realized depends on the succession of cellular contexts a cell type encounters during its differentiation.

\subsection*{Tracing cell type lineage maps with single-cell transcriptomics}

The first step in the construction of CTLMs is obviously the identification of an organism’s different cell types. Some cell types can be identified simply by their distinctive morphological features, such as the axonal projections of neurons, striped appearance of striated muscle cells, the disc-like shape of erythrocytes, etc. But such morphological and functional descriptions might be misleading. For example, smooth muscles in vertebrates and striated muscles in drosophila are morphologically and functionally distinct, but evolutionarily and developmentally equivalent \cite{brunet2016evolutionary}. Hence, how these different cells gain the ability to perform their functions, and how they are related to other cell types can only be seen through their molecular-level descriptions and gene expression patterns \cite{arendt2016origin}.

Recent advances in techniques such as single-cell transcriptomics, allow us to identify cell types with unprecedented accuracy and detail. These catalogs of cell types on their own already produce insights into the functioning of an organism, for example, the presence of neurons indicates that the organism is capable of transmitting information across its body.
But in order to gain insights into the process of development, cell types need to be arranged into CTLMs \cite{tritschler2019concepts}. 

Several recent studies already report CTLMs reconstructed from single-cell transcriptomics data \cite{plass2018cell,siebert2019stem}. However, there still remain many technical challenges in interpreting this data \cite{lahnemann2020eleven}. Most notably, the current algorithmic methods used to infer lineage maps are biased to preferentially produce \hyperref[glossary]{tree}-like and \hyperref[glossary]{chain}-like topologies \cite{tritschler2019concepts}.

The deeply rooted idea that CTLMs resemble trees probably owes itself to the history of cell lineage maps: The first cell-lineage map of embryonic development ever constructed was that of \textit{C.elegans}, and this map is remarkably tree-like \cite{girard2007wormbook}. Additionally, another extensively-studied lineage map is that of human hematopoeisis, which is also tree-like \cite{pellin2019comprehensive}. 
Perhaps the "tree archetype" has persisted due the tendency of the human mind to extrapolate: cells of multicellular organisms typically divide by binary fission. Evidently then, the two resulting daughter cells may assume at most two distinct cell-fates. In addition, experimental studies of differentiation typically look at the conditions that lead a daughter of a stem cell towards one of two alternative cell fates \cite{guo2010resolution, treutlein2014reconstructing}. The inverse problem of how two different cell types can differentiate convergently to produce the same cell type is much less studied.

While it is true that cellular differentiation is a branching process, it is erroneous to conclude that development, by extension, must look like a binary tree. In the first place, branches are parts of all connected graphs excluding simple chains and elementary cycles. Moreover, apart from the famous examples of tree-like lineage maps, we now also have examples of non-treelike lineage maps: zebrafish development \cite{wagner2018single}, and hydra adult homeostasis \cite{siebert2019stem}.

As an aside, an interesting recent theoretical work looks at how tree-like spatial arrangement of cells could facilitate the origin of multicellularity through the specialization of reproductive germ cells and non-reproductive somatic cells \cite{yanni2020topological}. But even if we assume that the spatial arrangement of cells imposes constraints on the differentiation of incipient multicellular species, it does not exclude the possibility that further elaboration of this species could lead to non-tree-like differentiation trajectories. 

For all these reasons, it is essential that we deal with the biases in lineage map reconstruction procedures so that the resulting picture of development remains faithful to biological reality. Some studies overcome these biases by using single-cell transcriptomics alongside cellular barcoding \cite{schmidt2017quantitative, wagner2018single}, which allows unambiguous identification of cellular lineages \cite{kebschull2018cellular}. It is instructive to note the progress made in phylogenetics, another field where the idea of phylogenetic 'trees' is prevalent. Recent phylogenetic inference algorithms allow the inclusion of reticulate events like hybridization and horizontal gene transfer \cite{huson2006application}, which yield non-tree-like networks. These methods could illustrate how one could resolve biases towards tree-like topologies in cell lineage reconstruction algorithms.

\subsection*{Elaboration of organisms through cell type invention}
Traditionally, cell types are a concept relating to multicellular organisms, but unicellular organisms are also known to switch between cellular phenotypes. They do so either according to temporal programs (Fig.\ref{fig1}(A)), or in response to changes in their environment (Fig.\ref{fig1}(B)) \cite{sebe2017origin}. Cells in yeast colonies even show spatial organization of distinct cellular phenotypes \cite{varahan2019metabolic}. Moreover, the unicellular organisms that are closest to multicellular lineages possess homologs of many "multicellularity related" genes, for example \textit{Chlamydomonas reinhardtii}, the closest unicellular relative of the volvox, has a homolog of regA, the gene responsible for cell type differentiation in volvox \cite{konig2020genetic}. 

A plausible origin of multicellularity was a transformation from an environmental/temporal regulation of cellular phenotypes to a developmental/spatial one. For example, in volvox, regA expression causes differentiation of the reproductive cell to the somatic cell (Fig.\ref{fig1}C), while the homolog of regA in \textit{C. reinhardtii} responds to environmental stresses, and induces a similar switch from a reproductive to a non-reproductive cellular phenotype \cite{konig2020genetic}. A similar scenario played out in the case of the facultatively multicellular dictyostid amoebas; the signaling molecules responsible for differentiation in multicellular species are instead produced in response to cold stress in unicellular dictyostelids \cite{kin2021evolution}.

\begin{figure}[hbt!]
\centering
\includegraphics[width=1.0\columnwidth]{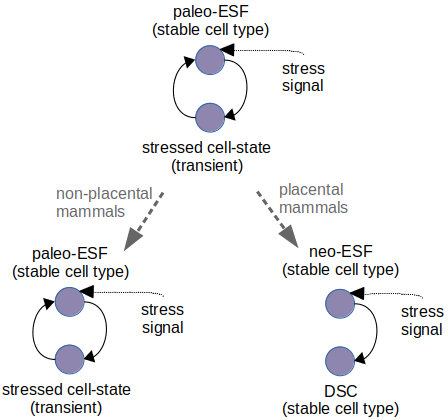}
\caption{Evolution of gene regulatory network in mammals leads to the invention of a new cell type. Blue circles represent cell-states, and bold arrows represent a change in cell-state. Dashed grey arrows represent evolutionary transitions. In the ancestor of marsupials (non-placental) and placental mammals, paleo-ESF cells responded to stress signals by elevating the expression of genes associated with stress-responses and apoptosis. The stressed cell state then relaxes back into the normal paleo-ESF cell. But in placental mammals, a rewiring of the regulatory network led to the invention of two new cell types: the neo-ESF and the DSC cells. The neo-ESF cell, upon receiving stress signals, differentiates into the DSC cell instead of expressing stress response genes \cite{erkenbrack2018mammalian}.}
\label{fig2}
\end{figure}

Multicellularity has evolved multiple times in the history of life, but multicellular organisms have expanded their cell types and evolved complexity only in six lineages: once in animals, once in plants, twice in algae, and twice in fungi \cite{knoll2011multiple}. The mechanism underlying the origin of multicellularity, which involved internalization of environmental cues, also plays a role in the invention of new cell types in already multicellular organisms. For example, the invention of DSC cells in placental mammals involved the internalization of stress signals into a cue for the differentiation of ESF cell. It also required rewiring of gene regulation such that the stressed ESF cell-state, which would normally relax back to its non-stressed state, now differentiates into a new cell type, the DSC (Fig.\ref{fig2}).

\subsection*{Cell type evolution encoded in gene regulation and reflected in developmental programs}
In \cite{arendt2016origin}, the authors anticipate changes in the genetic architecture that could lead to the invention of new cell types. In order to do this, they first emphasize the distinction between the function a cell type assumes in an organism, and the gene regulation that ensures its stability. In other words, cell types are stable states of \hyperref[glossary]{gene regulatory networks} (GRN) (see \cite{gershenson2004introduction} for a treatment of GRNs as random Boolean networks). 

Now, a multicellular organism can only contain a subset of the stable states prescribed by its GRN: these are the cell types which are accessed by its developmental program. GRNs also specify the dynamics through which transient cell-states transform into stable cell types~\cite{rand2021geometry}. We can imagine a range of molecular mechanisms, such as mutations--- affecting the activity of gene products, or altering molecular interactions, etc.---that could rewire the GRN. And there are three ways in which such rewiring can lead to cell type "invention":
(i) A frequently encountered transient cell-state shifts the stable cell type it maps to (this mode probably led to the invention of DSC cells, see Fig.\ref{fig2}); (ii) A transient cell-state becomes stable; (iii) The GRN expands by adding new genes, such as transcription factors, through gene duplication and divergence, horizontal transfer, etc. \cite{knoll2011multiple}.

The prevalent modes of cell type invention could be different in different multicellular lineages. We expect that developmental trajectories, and therefore CTLMs, reflect these frequently occurring modes. At the same time, we are aware that the phylogenetic trajectories of cell types and developmental trajectories need not be isomorphic \cite{arendt2016origin}. That is, although we expect modes of cell type invention to impose constraints on the topology of CTLMs, the exact form of this constraint is not obvious.

Moreover, development is also plastic and trajectories leading to particular cell types can shift. For example, in the ancestral multicellular dictyostelids, stalk-cells likely differentiated from pre-spore cells, whereas in the more recent group-4 dictyostelids, stalk cells differentiate from the developmentally distinct pre-stalk cells  \cite{kin2021evolution}. The potential for this plasticity can also be seen in our own cells, where expression of a single transcription factor, MYOD, can switch a fibroblast into a skeletal muscle cell \cite{arendt2016origin}. Over time, this plasticity can reconfigure CTLMs and potentially erase the signatures of the history of cell type invention.

From all this we see that the question of developmental evolution can be broken down into two parts: (1) the manner in which evolutionary trajectories of cell types shape developmental trajectories, and (2) the extent to which this signature is preserved despite the plasticity of development. In \cite{yuan2020alignment}, the authors demonstrate the relationship between cell type evolution and CTLM topology; they show that the evolutionary closeness of cell types can be identified through a comparison of CTLM subgraphs rooted at these cell types. We suggest here that beyond the detection of evolutionary closeness, global patterns of cell type evolution within multicellular lineages can be revealed through statistical analyses of CTLM topologies.

\section*{Anticipating modes of cell type invention from a generative model}

As a concrete example for how mathematical models can bridge the regulatory architecture of development and CTLM topologies, we examine here a minimal generative model~\cite{mani2021comprehensive}. Our model incorporates three fundamental features of development:
\begin{enumerate}
    \item There exist multiple stable cell types and transient cell-states that reliably map to specific stable cell types.
    \item The fate of any cell depends on its cellular context due to cell-cell signaling. For example, during \textit{Drosophila} oogenesis, cellular exchange of growth regulating proteins among the oocyte and surrounding nurse cells sets up a spatial coordinate system which determines the subsequent growth behavior of nurse cells \cite{doherty2021coupled}.
    \item Since development of a multicellular organism begins with a single cell, mechanisms for symmetry breaking, such as asymmetric cell division and cell polarization, that can create an interactive field of cells, are essential.
\end{enumerate}

\subsection*{A generative model of development}
\begin{figure*}[hbt!]
\centering
\includegraphics[width=0.6\textwidth]{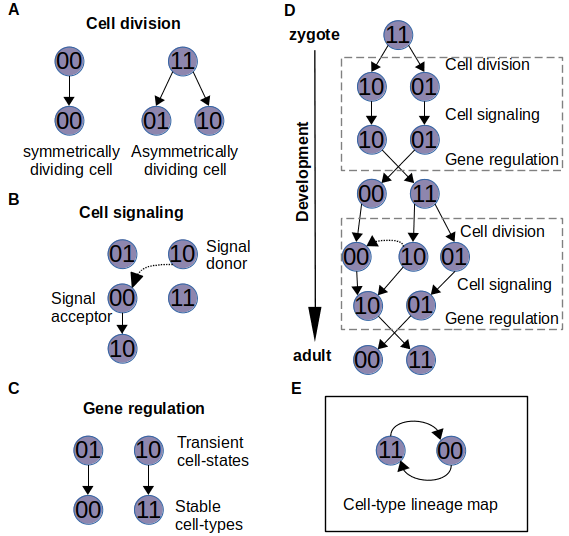}
\caption{Generative model of biological development. Blue circles represent cell-states, numbers written inside them represent the identity of cell-states; 0 represents the absence and 1 represents the presence of a cell-state determinant. Solid arrows represent change in cell-state, and dashed arrows represent exchange of signaling molecules. (A,B,C) represent the regulatory architecture of an organism with N = 2 determinants. (A) Asymmetric cell division: cell types in the model can produce daughter cells which are not identical. (B) Cell signaling: certain cell-state determinants can act as signals and are secreted by donor cells and received by specific acceptor cells. In this example, the first determinant is a signal. The state of the acceptor cell reflects signal reception by switching the state of signal determinant to ‘1’. (C) Gene regulation: certain cell-states are stable cell types, and others are transient cell-states which map to the stable cell types. (D) Scheme of development in the model: the \textit{zygote} undergoes repeated rounds of asymmetric cell-division, cell-cell signaling, and gene regulation according to the rules outlined in (A,B,C). The process is iterated until the resulting set of cell types repeats itself; this set of cell types forms the adult. (E) Cell type lineage map (CTLM) of adult homeostasis: in this example, the two cell types constitute the adult produced by the developmental program sketched in (D). The arrows represent differentiation.}
\label{fig3}

\end{figure*}

We capture these features of development by decomposing the regulatory architecture of development into three universal components: asymmetric cell division, cell signaling and gene regulation. For simplicity, we encode these three components as Boolean logical functions. This  coarse-grained representation does not depend on details specific to any particular organism, and can therefore describe a wide variety of organisms. The simplicity allows us to sample millions of developmental programs as combinations of rules for cell division, signaling and gene regulation (Fig.\ref{fig3}).

In the model, "genes" of organisms encode for \textit{cell-state determinants}, which are interacting sets of transcription factors that combinatorially determine the identity of the cell \cite{arendt2016origin}. For an organism with $N$ genes, cell-states are defined as binary strings, where ‘1’ denotes the presence and ‘0’ denotes the absence of a cell-state determinant. Cells in the model divide asymmetrically, to produce unequal daughter cells (Fig.\ref{fig3}A). Of the $2^N$ possible cell-states, a few are assigned the status of stable cell types, while other cell-states are transient and map deterministically to one of the stable cell types (Fig.\ref{fig3}C). These assignments represent gene regulation in the model. Finally, certain cell-state determinants can be exchanged as signals among specific donor and receiver cells (Fig.\ref{fig3}B). This allows cells to interact with their cellular context. 

We model development as an unfolding process, which begins with a randomly picked initial cell type, the \textit{zygote}, and proceeds through rounds of asymmetric cell-division, cell-cell signaling, and gene regulation (Fig.\ref{fig3}D). This sequence of operations is repeated until the resulting set of cell types repeats itself. This final set of cell types is a steady state of the model, which we call the adult. Within an organism in the model, the regulatory architecture is fixed throughout the development process. This algorithm can also model \textit{regeneration}; here, instead of starting with the zygote, we represent an \textit{injury} as a loss of a subset of adult cell types, and initialize the developmental program with the remaining adult cell types.

In the model, a cell type \textit{x} differentiates into a cell type \textit{y} if one of the daughter cells of \textit{x} gives rise to \textit{y} after one round of signaling and gene regulation. In this way, we can construct CTLMs, which are graphs whose nodes represent cell types and directed edges represent differentiation (Fig.\ref{fig3}E). A CTLM generated by the model can represent various stages of development, 
\begin{itemize}
    \item Embryonic development: when the nodes of the graph include all the cell types encountered starting from the initial zygotic cell to the final adult.
    \item Adult homeostasis: if the nodes of the graph only include the adult cell types (as in Fig.\ref{fig3}E).
    \item Regeneration: if the nodes of the graph include cell types produced during regeneration.
\end{itemize}

 The model is `generative' because we can randomly draw a large number of developmental programs and their corresponding CTLMs, typically a few millions, and statistically analyze their features. Even such a large set remains minute compared to the overall number of potential developmental programs. However, the sampling of the set provides reliable statistics, and thereby overcomes the `curse of dimensionality'~\cite{eckmann2021} that follows from Cantor's theorem.

\subsection*{Insights from the generative model}
In \cite{mani2021comprehensive}, we use the model described above to generate developmental programs in an unbiased manner, and look at the homeostatic adult CTLMs it produces. This is the distribution of cell-lineage map topologies we should expect, for instance, in multicellular lineages where developmental plasticity has erased all traces of the mode of cell type invention.

\begin{figure*}[hbt!]
\centering
\includegraphics[width=0.85\textwidth]{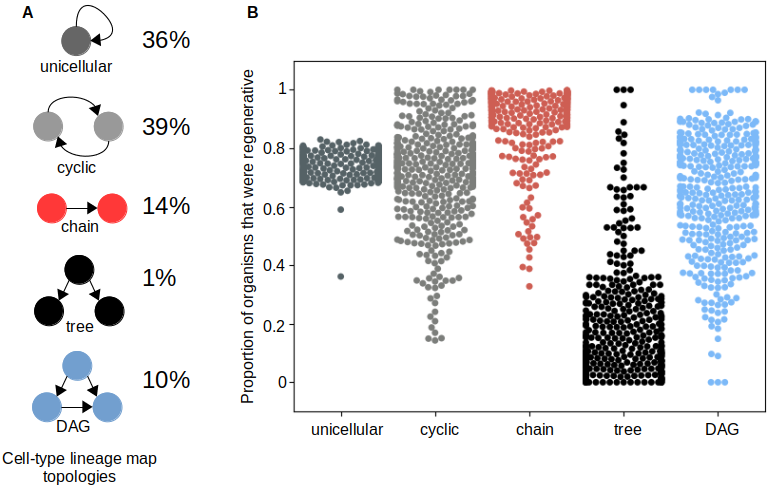}
\caption{Prevalence of topologies and their regenerative capacities. (A) Graph topologies of cell type lineage maps (CTLMs). The numbers beside the graphs indicate their prevalence in our data. (B) Distribution of regenerative capacities of organisms with different CTLM topologies is represented by the distribution of points in the swarms.}
\label{fig4}

\end{figure*}

The CTLMs generated by the model are classified according to their graph topologies: unicellular, \hyperref[glossary]{cyclic}, chain, tree, and \hyperref[glossary]{\textit{directed acyclic graphs}} (DAG) (Fig.\ref{fig4}A). Here, DAG specifically refers to acyclic graphs that are not tree-like and possess edges cross-linking different branches.  Since differentiation in multicellular organisms is generally assumed to be irreversible, acyclic CTLMs, chain, tree and DAG, are more likely to be biologically relevant. Counter to expectations, our results indicate that tree-like graphs were extremely rare (~1\% of all lineage maps). The most common acyclic lineage maps were simple 2-node chains, which resemble volvox (Fig.\ref{fig1}C). Among the more complex acyclic graphs with more nodes, DAGs were the most common (as is the case for hydra, in Fig.\ref{fig1}D). Thus, our results suggest that CTLMs representing the adult stage of complex multicellular organisms is very likely to be a DAG. 

In reality, we expect that not all traces of the mode of cell type evolution have been wiped out due to developmental plasticity. That is, some biases and patterns should persist in the regulatory architecture of development, and thereby could leave a signature on the cell-lineage map topologies. In principle, such patterned developmental programs can be described by the generative framework, which can be used to also survey and predict the distribution of CTLM topologies for distinct modes of cell type invention. Such a study could be used for comparison with experimentally obtained cell lineage maps, allowing us to assess the modes of cell type invention prevalent in different multicellular lineages.

The topologies of generated CTLMs also indicate the ability of organisms to regenerate. The regenerative capacity of organisms was computed by separating single cell types from the adult and testing whether they were able to regenerate the complete adult. In other words, we tested whether the adult organisms of the model contained \textit{pluripotent cells}. While we found organisms with chain-like and DAG-like lineage maps to be remarkably regenerative, those with tree-like CTLMs turned out to be the least regenerative (Fig.\ref{fig4}B). This illustrates that topologies of CTLMs---in addition to being a summary of developmental events, and indicative of the evolution of developmental programs---can also hold functional information, in this case, the ability to regenerate.

\section*{Conclusions and outlook}

The evolution of multicellular complexity is synonymous with the evolution of cell types in multicellular lineages. Adding new cell types to an organism involves rewiring of its gene regulatory network. We have outlined here three modes through which such rewiring of the regulatory architecture can lead to cell type invention.

The gene regulatory network of an organism prescribes not only its cell types, but also developmental routes to reach these cell types. Thus, the prevalent modes of cell type invention are also likely to be reflected in its CTLMs. 
Mathematical models of development, like ours, can be used to map the regulatory architecture of development to CTLMs. These can be useful in anticipating the properties of CTLMs resulting from developmental programs that have been sculpted by different modes of cell type evolution. 

We used the model to survey millions of developmental programs, unbiased towards any particular mode of cell type evolution. The survey produced a characteristic distribution of cell lineage map topologies: tree-like lineage maps were extremely rare, and complex multicellular lineage maps were more likely to be represented by directed acyclic graphs (DAGs).
We suggest that a combination of modeling approaches that predict topologies of cell lineage maps, and surveys of cell lineage maps of real organisms could uncover the patterns of developmental evolution in the various multicellular lineages.
The present results demonstrate that minimal coarse-grained models of development could serve as a complementary approach to detailed molecular models: while detailed models describe the implementation of development in a particular organism, coarse-grained models can be used to scan a huge space of programs and produce general conclusions about developmental processes. Importantly, mathematical models allow simplified conceptions of complicated biological processes, and produce experimentally testable predictions about core features of the functioning, origin and evolution of these processes.

\section*{Glossary}\label{glossary}
The following are formal definitions of a few central terms, which are mostly familiar on an intuitive level. \\

\noindent\textbullet~~\textbf{Multicellular complexity}: A concept that is roughly equivalent to the number of distinct cell types an organism possesses. Simple multicellular organisms posses 2 cell types, usually a germ cell and a somatic cell, whereas complex multicellular organisms have elaborate bodies and can contain hundreds of cell types.

\noindent\textbullet~ \textbf{Developmental program}: The set of instructions that can be \textbf{(a)} used by the single celled zygote to form a complete adult, or \textbf{(b)} for the maintenance of adult body, or \textbf{(c)} for regeneration of the body post injury.

\noindent\textbullet~ \textbf{Cell type lineage map (CTLM)}:  A graph whose nodes are the different cell types of an organism, and a directed edge between two nodes indicates that one cell type differentiates into the other during the course of the organism's development.
    
\noindent\textbullet~~\textbf{Cellular context}: The set of cell types that co-occurs with a given cell type in the developing body. Cells interact with their cellular context through signaling, and this interaction regulates their differentiation trajectory.
    
\noindent\textbullet~~\textbf{Graph topology}: The manner in which the nodes and edges of a graph are arranged. The topological properties of graphs includes the length of its shortest path, the degree distribution of its nodes, presence of node clusters, loops, etc. Certain graph topologies are particularly recognizable: such as cycles, chains and trees.
    
\noindent\textbullet~~\textbf{Cyclic graph}: A graph which contains at least one cyclic path, \textit{i.e.}, a path which begins and ends at the same node. Presence of cycles in CTLMs indicate programmed reversibility of cellular differentiation.
    
\noindent\textbullet~~\textbf{Chain}: A connected linear graph which contains exactly one directed path between a starting node and end node. A 2-node chain CTLM represent simple multicellular organisms such as volvox. 
    
\noindent\textbullet~~\textbf{Tree}: Any connected acyclic graph with $n$ nodes and $n-1$ edges, where $n$ is an integer greater than 2. Such graphs are characterized by paths that look like branches of a tree. In CTLMs with tree-like topologies, there is exactly one developmental route that can access any cell type. 
    
\noindent\textbullet~~\textbf{Directed Acyclic Graph (DAG)}: A connected graph which contains no cycles. While chains and trees are also DAGs, in general, DAGs contain edges that link its different branches. In a CTLM, these links represent multiple developmental routes that converge on the same cell type. 
    
\noindent\textbullet~~\textbf{Clonal multicellularity}: The form of multicellularity where the body of the organism grows by the repeated division of a single initial cell. The body is necessarily composed of clonally related cells. For example, animals and land plants. 
    
\noindent\textbullet~~\textbf{ Aggregative multicellularity}: The form of multicellularity where the body of the organism grows by the aggregation of cells of the same species. Therefore the organism need not be composed of clonally related cells. For example, spores of dictyostelids are aggregatively multicellular. 
    
\noindent\textbullet~~\textbf{Gene Regulatory Network (GRN)}: GRNs represent genetic interactions that regulate the activity of genes, or gene products. Mathematically, GRNs can be represented as graphs whose nodes represent genes and the presence of an edge between two nodes indicates that one of the genes regulates the activity of the other.

\section*{Author Contributions}

Authors designed research, wrote and edited the manuscript.

\section*{Acknowledgments}

This work was supported by the Institute for Basic Science, Project Code IBS-R020. We thank Luca Peliti, Albert Libchaber and Mukund Thattai for useful discussions about the model.

\bibliography{sample}

\end{document}